\documentstyle[12pt]{article}

\newcommand{\be}{\begin{equation}}
\newcommand{\ee}{\end{equation}}
\newcommand{\bdis}{\begin{displaymath}}
\newcommand{\edis}{\end{displaymath}}

\begin{document}
\centerline{\LARGE Cosmological Principle and the debate about
Large Scale Structures distribution}
\centerline{\LARGE L.Pietronero$^{1}$ and F. Sylos Labini$^{1,2}$}
\centerline{\footnotesize ($^1$) Dipartimento di Fisica, Universit\'a di Roma
``La Sapienza'', P.le A.Moro 2,Rome, Italy}
\centerline{\footnotesize ($^2$) Dipartimento di Fisica, Universit\'{a} di
Bologna,Italy}

\section{Introduction}

The basic hypothesis of a post-Copernican Cosmological theory is that
{\em all the points} of the Universe have to be essentially equivalent:
this hypothesis is
required in order to avoid any privileged {\em observer}.
This assumption has been implemented
by Einstein in the so-called
 Cosmological Principles (CP): {\em all the positions}
in the Universe have to be essentially equivalent,
so that the Universe is homogeneous.
This situation implies also the condition of spherical
symmetry about every point, so that the Universe in also Isotropic.
There is a hidden assumption in the formulation of the CP in regard to
the hypothesis
that all the points are equivalent
on which we will return later.

 The condition that all the occupied points
are statistically equivalent with
respect to their environment correspond to the property of
of Local Isotropy. It is generally believed that the
Universe cannot be isotropic about every
point without being also homogeneous [1].
 Actually Local Isotropy does not necessarily implies homogeneity;
in fact a topology theorem states that homogeneity
is implied by the condition of local isotropy
 together with {\em the assumption
of the analyticity or regularity} for the distribution of matter.

Up to the seventies analyticity was an obvious implicit assumption
in any physical  problem. Recently however we have learned about intrinsically
irregular structures for which analyticity should be considered as a
property to be tested with appropriate analysis of experiment
([2],[3]).
\bigskip

The current idea is that in the observable Large
 Scale Structure distribution,
isotropy and homogeneity do not apply to the Universe
in detail but only to a "smeared- out " Universe
 averaged over regions of order
$\:\lambda_{0}$.
One of the main problems
of observational cosmology is therefore
the  identification of $\:\lambda_{0}$.
\smallskip

Several recent galaxy redshift
surveys such as CfA 1 [4],
CfA 2 [5], [6], SSRS1 [7], SSRS2 [4]
Perseus Pisces [8]
and also pencil beams surveys [9],
have uncovered remarkable structures such as
 filaments, sheets, superclusters
and voids. These galaxy surveys
now probes scales up to $\: 200-300 h^{-1} Mpc$ and show that
the Large Scale Structures are relatively common.

One of the most important issues
raised by these catalogues is that the scale of {\em the
largest inhomogeneities} is comparable with the
extent of the surveys and that the largest known
structures are limited by {\em the boundaries of
 the survey} in which they are detected.
Da Costa et al. [6] emphasize that
both CfA2 and SSRS2 contain voids with
diameters as large as $\:50 h^{-1} Mpc$
 and that the Southern Survey contains the
Southern Wall, similar to the Great Wall
 of the northern CfA: the general nature
 of the LSS distribution in the Southern
Survey is similar to the Northern one.
Moreover Da Costa et al.[6] examining the
normalized density fluctuations in the CfA2
and SSRS2 surveys, concluded that there
are fluctuations in shells of  $\:100 h^{-1} Mpc$
and that the combined sample is not homogenous.

Finally it is remarkable to note that
Tully et al.[10] analyzing the combined Abell and ACO clusters
catalogues provide that there are evidences of
structures on scale of $\:450 h^{-1} Mpc$
lying in the plane of the Local Supercluster.
\bigskip

This observational situation appears therefore highly
problematic with respect to the identification of the length
$\:\lambda_{0}$ above which the distribution
should be smooth and essentially structureless.

\section{Observations and analysis}

Coleman and Pietronero [11], analyzing
with the methods of modern statistical mechanics the
Cfa 1 [4] for galaxies and the Abell
catalogue [12] for clusters, find that these samples
shows power law (fractal)
correlations up to the
sample limit without any tendency towards
homogenization. In particular they find that the number-number
correlation function (or other related quantities):
\be
G(r) = <n(\vec{r_{0}})n(\vec{r_{0}}+\vec{r})>
\ee
has power law behaviour:
\be
G(r) \sim A r^{-\gamma}
\ee
with $\:\gamma \sim 1.6-1.8$.
If the distribution would really  become homogenous at some length-scale
$\:\lambda_{0}$ within the sample, one should instead observe a power law
followed, for $\:r>\lambda_{0}$, by {\em a well defined flat behaviour}.

The usual analysis consists in computing the
two-point correlation function [13]:
\be
\xi(r) = \frac{<n(\vec{r_{0}})n(\vec{r_{0}}+ \vec{r})>}{<n>^{2}}-1
\ee
where $\:n(r)$ is the galaxy number density and $\:<n>$
is the average over the entire sample. At small scales one observes a power
law behaviour [14]:
\be
\xi(r) \sim A r^{-\gamma}
\ee
where $\:\gamma \sim 1.7-1.8$ both for galaxies and clusters catalogues.
By the way we note that comparing eq.(1) and (3) it results that
 in the case of self-similar
distribution, $\:\xi(r)$
is a power law only when $\:\xi(r) \gg 1$.
The so-called "correlation
length " in the standard approach
is defined by the relation:
\be
\xi(\vec{r_{0}})= 1
\ee

For galaxy distribution one obtains $\:r_{0}^{G} \sim 5 Mpc$,
while for clusters  $\:r_{0}^{C} \sim 25 Mpc$.
{}From this analysis it follows that
the amplitude of $\:\xi(r)$ is different for
galaxies and clusters and
the clusters are more correlated
than galaxies. This observation
has been one of the fundamental points
of the so-called biased galaxy formation model [15].
\smallskip

The basic point of the debate about galaxies and clusters distributions
and correlations [11] is
that the analysis with the function $\:\xi(r)$ is meaningful
only if the quantity $\:<n>$, that enters
explicitly in the definition of $\:\xi(r)$ but
not in $\:G(r)$, is a well defined quantity,
i.e. it does not
depend on the
sample size. If, on the other hand,
 the distribution has correlations
extending beyond the sample limit,
the value of $\:<n>$ will be a
direct function of the sample size. In this case the
value of $\:r_{0}$ obtained via
eq.(5) does not measure  the real correlation
length towards homogeneity, but it correspond
just to a fraction of the sample size and it has no physical meaning
with respect to the real correlation properties of galaxies
and clusters. In other word the amplitudes of the correlation
function of galaxies and clusters are not physical quantities but
are related to the sample size.

Often the concept of {\em "fair sample"}
 is confused with "homogeneous sample"
while these are two separate concepts.
A sample is statistically fair if it is possible
to extract from it information that is {\em statistically meaningful}.
Whether it is homogeneous or not is a property that can
be tested and it is independent from statistical fairness.
\bigskip

For CfA 1 and Abell catalog Coleman
and Pietronero [11] have obtained the following results:

- the CfA 1 catalogue is statistically a "fair sample" up to $\:20 h^{-1}Mpc$:

- The correlation properties in the Cfa 1
show a well defined power law (fractal)
behaviour up to the sample limits without
 any tendency towards homogenization.
\smallskip

- The Abell catalogue shows a well defined power law behaviour without
any tendency
towards homogeneity
within the limits of statistical validity of the sample ($\:60-80 h^{-1}Mpc$)
\smallskip

Now we can clarify that the mismatch between galaxy and clusters correlation
is just due to an inappropriate
mathematical analysis. The correlation lengths
$\:r_{0}^{G} \sim 5 Mpc$ for galaxies
 $\:r_{0}^{C} \sim 25 Mpc$ for clusters
are a finite portion of the depth of the
galaxies ($\:R_{G}$) and clusters
($\:R_{C}$) catalogues. For Abell and Cfa 1 one
finds that:
\be
\frac{r_{0}^{C}}{r_{0}^{G}} \sim \frac{R_{C}}{R_{G}}\sim 5
\ee

Therefore the correlation
of clusters appear to be the continuation of galaxy correlations to
larger scales {\em (Fig.1)} and
the two samples simply refer to different observations of the system - which
has fractal correlation up to the present observational limit.
This amplitude rescale is in perfect agreement with the self-similar
behaviour of galaxies and clusters distributions [11].
\bigskip

The fact that the amplitude of $\:\xi(r)$, and therefore
also $\:r_{0}$, are larger for clusters than for galaxies has
given rise to the statement that "clusters correlate to larger
distance than galaxies". This proposition has been extended
also among galaxies of different luminosity and the idea is that
more luminous galaxies correlate to larger distances
than less luminous ones.
Since clusters are made of galaxies and voids are
empty for both galaxies and clusters
such statement is technically inconceivable.
To clarify this point we recall that $\:r_{0}$
is a length related to
the sample depth and has no physical meaning.
Sample with more luminous galaxies are typically deeper
than those with less luminous galaxies and this is the real
origin of the effect.
Therefore the {\em "galaxy-cluster
mismatch"} and the {\em "clustering richness
relation"} are problems that arise only from unreal quantities
like the amplitude of $\:\xi(r)$ and $\:r_{0}$. Within a
correct analysis these problems automatically
 disappear and cluster correlations
correspond naturally to galaxy correlations.
The attempt of rescaling the two correlations via lower
cut-off off length $\:\lambda$ proposed by
Szalay and Schramm [16], looses now its basic motivation:
moreover this would not be the correct rescaling for
self-similar system [11].
\bigskip

\begin{figure}
\vspace{8cm}
\caption{Ideal catalogue of galaxies and clusters.
The correlation
of clusters appear to be the continuation of galaxy correlation to
larger scales: the points are galaxies and the groups of galaxies are
identify as clusters. The depth of the clusters catalogues
is deeper than that
of galaxies catalogues because clusters are brighter than galaxies.}
\end{figure}

In addition to the homogeneity scale $\:\lambda_{0}$,
 it is interesting to
identify the isotropy scale $\:\lambda_{I}$,
i.e. the scale over which
the statistical isotropy of the galaxy
(or cluster) distribution has been reached.
The evidence of dipole saturation in galaxies and clusters
catalogues [17],
together with a monotone growth of the monopole,
shows that the isotropy scale has been reached,
i.e. that $\:\lambda_{I}$ is less than the catalogues depth.
Since a fractal structure has usual the property of
 local isotropy,
it is fully compatible with the evidence of dipole
saturation with depth as well as an homogeneous
distribution, as shown in {\em Fig.2} [18].
The isotropy scale cannot be simply related to
the homogeneity scale unless the distribution is smooth
(analytic), but we have seen that it is certainly not the case.
\begin{figure}
\vspace{8cm}
\caption{Monople and dipole modulus
behaviour with sample depth for a fractal with $\:D=1.4$:
the evidence of dipole saturation is
a proof of Local Isotropy and not of homogeneity}
\end{figure}
\bigskip

Very recently we have also
 analyzed the ESO Slice Project (ESP) [19]
that is a galaxy redshift survey over a strip
of $\:22^{\circ} * 1^{\circ}$ in the South Galactic Pole region
and with a limiting magnitude of $\:b_{J} \le 19.4$. The total number of
objects is of the order of $\:4000$.
Due to the small solid angle covered by this survey it is not
possible to compute the number-number correlation function, while it
is possible to study the {\em number-redshift}  relation $\:N(z)$,
that gives the total number of galaxy
within a spherical symmetric volume of redshift $\:z$. This
is one of the crucial test for world models [20]
and gives unambiguously, and
free of any a priori assumption,
the properties of the galaxy number density.Our preliminary analysis
shows an indication that in the ESP survey
the distribution of galaxies is not homogenous
up to the sample depth ($\:\sim 500-700 h^{-1} Mpc$),
and that it shows power law
(fractal) correlations up to the sample limits;
the exponent of the fractal distribution is about 2.
\bigskip

The pictures that emerges with an analysis
of galaxy and cluster redshift surveys, that does not
imply any a priori assumption is quite different from the standard
one: the larger scale distribution of galaxies and clusters
shows well defined fractal properties
without any tendency towards
homogeneity {\em up to the present observational limits}.

\section{Properties of  a fractal distribution of matter}

A non-analytical distribution can be statistically isotropic above some
scale $\:\lambda_{I}$.
This means that all the points are statistically equivalent
with respect to their environment.
 A non-analytical  (fractal) distribution
breaks the symmetry between occupied and
 empty points: the translation invariance
is not satisfied any more even if the
distribution shows on average
 spherical symmetry around
every point. This means that each point
of the fractal structure is statistically identical
(in the sense of local isotropy)
to any other point of the structure. However in the middle
of a void the environmental properties
are different than for the points of the
 structure.

The cosmological principle can therefore
be maintained in term of Local Isotropy
in perfect agreement with the basic hypothesis of having no
privileged points [2], see also [21].
\bigskip

The absence of analyticity has important consequences.
First of all irregularity is intrinsic at all
scales and the structure never becomes
smooth (analytical). For this reason
the global averages like number density
or, as we have shown in the previous section, the amplitude of $\:\xi(r)$
are related to the sampled volume.
 In particular we stress that
the normalized density fluctuations $\:\delta N/N$ at scale of $\:8 h^{-1}Mpc$
suffers of the same problem of $\:r_{0}$
 and its value is the counterpart of the of
 the finite size of the sample and
has no direct physical meaning. Hence
one cannot identify the
scale at which the fluctuations
are small respect to the average
as the scale of the transition from non linear to
non-linear behaviour: this scale is a
consequence of an incorrect mathematical analysis
and a fractal structure is "non-linear" at all scales.
However considering a finite portion of a fractal structure,
one can always find a scale at which
$\:\delta N/N \ll 1$, but this is just
due the finite size of the system rather than to an intrinsic
physical properties. The distance at which
$\:\delta N/N =1 $ will scale with the sample depth,
as $\:r_{0}$ scales [22],
so it is not an evidence of homogeneity, or
any other ral change of nature of the distribution.

\section{The multifractal mass distribution}

Coleman \& Pietronero [11] have performed a multifractal analysis
of the Cfa1 for the full matter distribution,
by including also the galaxy masses: the result
shows indeed well defined multifractal properties.
This fact has a number of consequences
on various morphological properties. Here we discuss
the main one, i.e the correlation between space
and mass distributions.
\smallskip

We have seen that a basic characteristic of the observable
distribution of galaxies is the existence of LSS having
fractal properties at least up to some length that
increases with each new catalogue.
A second important observational feature is the galaxy  mass function:
this function determines the probability of having a mass in the
range $\:M$ to $\:M+dM$ for unit volume,
and can be described by the Press-Schecheter
function that has a power law behaviour in the limit of low mass,
followed by an exponential cut-off [23]:
\be
n(M)dM \sim M^{\delta-2} exp(-(M/M^{*})^{2\delta})dM
\ee
with $\:\delta \sim 0.2$.
\smallskip
These two observational evidences can be linked together by the concept
of multifractal (MF) that gives
an unified picture of the mass and spatial distributions.
A MF distribution describes systems with
local properties
of self-similarity [24]
and it is characterised by a continuous set of exponent
$\:f(\alpha)$.
This distribution implies a strong correlation between spatial
and mass distributions so that the number of object with
mass $\:M$ for unit volume $\:\nu(M,\vec{r})$, is a complex function
of space and mass and is not simply separable in a density
function$\:D(\vec{r})$ multiplied by a mass function
$\:n(M)$, as usually done.
\bigskip

It can be shown [25]
that the mass function of
a multifractal in a well defined volume,
 has a Press-Schechter shape with the
exponent $\:\delta$ that depends on the shape of $\:f(\alpha)$.
Moreover the fractal dimension of the support
is $\:D(0)=f(\alpha_{s})=3-\gamma$ (eq.(2)).
 Hence with the knowledge of the
whole $\:f(\alpha)$ spectrum one obtains information on the
space  correlation and on the mass function.
\smallskip

The evidence that galaxian luminosity and location in space are not
independent and that, on the contrary,
are strongly correlated have important
consequences in all the luminosity properties of visible matter
and in relation to
the methods that are usually used to study these properties
[11], [25].
\smallskip

{}From a theoretical point of view the
problem is therefore to identify
the dynamical processes that leads to such a multifractal
distribution, as discussed in [26] on this issue.

\section{Conclusion}

The new experimental picture that comes out from the redshift
survey is that the distribution of visible matter is fractal
and multifractal up to the present observational limits
(see {\em Fig.3}). New deep  redshift surveys
as the ESP survey [20],
but also pencil beam surveys [9],
suggest that the cut-off towards homogeneity
is deeper than the survey limit $\:\sim 500-600 h^{-1} Mpc$.
\begin{figure}
\vspace{8cm}
\caption{The large scale
 distribution of visible matter in the universe according
to the usual picture (above) compared to the one that aries from our new
studies (below).}
\end{figure}
\bigskip

We have shown that a fractal (and multifractal)
distribution is, of course, not homogenous and can be
locally isotropic so that it holds the basic hypothesis that
all the points of the Universe are essentially equivalent and  that
there are not privileged observer.
The Cosmological Principle can be naturally
generalized
to the case of non-analytical distribution
that breaks the symmetry between occupied and empty
points.
\bigskip

The relation of the new picture to the metric and Einstein's equation
depends crucially on the eventual properties of the dark matter.
If this would turn out to be usual
and distributed homogeneously there
is basically no problem with the predominant description of the
metric, the Big-Bang model, etc.
In the opposite situation the problem becomes very hard
and it has to be reconsidered from the
beginning. In any case some fundamental aspects
of the currently theories of galaxy formation, such
as the biased galaxy formation model and related theories,
have fundamental problems with the new picture that we
have briefly described.
\smallskip

The usual discussion of biased galaxy formation
is implemented by arguing that before and after
the evolution of fluctuations one has two
different types of density fluctuations
both however within an analytic Gaussian framework.
Thus, neither
 long range correlation are present nor
the power law behaviour. The self-similarity and the
 non-analyticity correspond to a breakdown of the Central Limit Theorem
which is instead
the necessary cornerstone of gaussian process.
In addition a power law's amplitude has virtually no
meaning while the exponent is the crucial quantity, so that one
has to approach the problem with a fundamentally
different theoretical
framework (see for example [3])
\bigskip

\section*{Acknowledgments}
We thank Dr. G. Vettolani for
useful discussions and for giving us the opportunity
of analyzing the ESP survey.

\section*{References}
\begin{itemize}
\item [\mbox{[1]}] Weinberg, S., 1972 Gravitation and Cosmology (Weley; New
York)
\item [\mbox{[2]}] Mandelbrot, B., 1982 The fractal geometry of nature,
Freeman New York
\item [\mbox{[3]}] Pietronero, L., \& Tosatti, E. (eds) 1986, Fractals in
Physics
\item [\mbox{[4]}] Huchra J., Davis,M.,Latham,D. and Tonry,J. 1983 ApJ Supp.
52,89
\item [\mbox{[5]}] de Lapparent, V., Huchra, J., Geller, M. 1988 ApJ 332, 44
\item [\mbox{[6]}] Da Costa, N.,et al., 1994 ApJ 424,L1
\item [\mbox{[7]}] Da Costa, N.,et al., 1988 ApJ 327,544
\item [\mbox{[8]}] Giovanelli, R., \& Haynes, M., 1993 AJ,105,1271
\item [\mbox{[9]}] Broadhurst, T.J., Ellis, R.S.,
Koo, D.C., Szalay,A.S., 1990 Nature 343,726
\item [\mbox{[10]}] Tully et al. 1992, ApJ 388,9
\item [\mbox{[11]}] Coleman, P.H. \& Pietronero, L.,1992 Phys.Rep. 231,311
\item [\mbox{[12]}] Abell,G.O., 1958, ApJ Supp. 3,211
\item [\mbox{[13]}] Peebles, P.J.E., 1980 The Large Scale Structure of the
Universe,
 Princeton University Press, New Jersey
\item [\mbox{[14]}] Davis,M. \& Peebles P.J.E. 1983, ApJ 267,465
\item [\mbox{[15]}] Kaiser, N., ApJ Lett.,284,L9-12
\item [\mbox{[16]}] Szalay A.S \& Schramm, N.D 1985 Nature 314,718
\item [\mbox{[17]}] Scaramella, R.,Vettolani, G., Zamorani, G. 1991 ApJ 376,L1
\item [\mbox{[18]}] Sylos Labini F., 1994 ApJ in press.
\item [\mbox{[19]}] Vettolani, G. et al. 1993, in the Proceedings
of Schloss Rindberg workshop.
\item [\mbox{[20]}] Sandage,A., 1988 Ann Rev A\&A, 1988 26,561
\item [\mbox{[21]}] Ribeiro,M.B., 1993 ApJ 415,469
\item [\mbox{[22]}] Coleman,P.H.,Pietronero,L., Sanders,R.H 1988 A\&A 200,L32
\item [\mbox{[23]}] Press W.H \& Schechter, P. 1974 ApJ 187, 452
\item [\mbox{[24]}] Paladin,G., \& Vulpiani, A. 1987, Phys.Rep.,156,147
\item [\mbox{[25]}] Pietronero,L. \& Sylos Labini,F. in preparation
\item [\mbox{[26]}] Sylos Labini, F. \& Pietronero L., on this issue

\end{itemize}

\end{document}